\documentclass[pdf,epjc3]{svjour3}  
\smartqed  
\RequirePackage{graphicx}
\RequirePackage{mathptmx}      
\RequirePackage[T1]{fontenc}
\RequirePackage{flushend}
\RequirePackage[numbers,sort&compress]{natbib}
\RequirePackage[colorlinks,citecolor=blue,urlcolor=blue,linkcolor=blue]{hyperref}
\usepackage{amsmath,amsfonts,amssymb,revsymb,dcolumn,epsfig,bm,xspace}
\usepackage[utf8]{inputenc}
\usepackage{threeparttable}
\usepackage{multirow}

\newcommand{\htt}{h_\mathrm{t}}
\newcommand{\hs}{h_\mathrm{s}}

\journalname{Eur. Phys. J. C}

\begin{document} 

\title{Reconstruction of energy conditions from observations and implications for extended theories of gravity}

\author{M. Penna-Lima\thanksref{e1,addr1,addr2}
\and 
S.D.P. Vitenti\thanksref{e2,addr3,addr4} 
\and 
M.E.S. Alves\thanksref{e3,addr5} 
\and 
J.C.N. de Araujo\thanksref{e4,addr6} 
\and
F.C. Carvalho\thanksref{e5,addr7} 
}                     
%
%
\thankstext{e1}{e-mail: pennalima@unb.br}
\thankstext{e2}{e-mail: sandro.vitenti@uclouvain.be}
\thankstext{e3}{e-mail: marcio.alves@unesp.br}
\thankstext{e4}{e-mail: jcarlos.dearaujo@inpe.br} 
\thankstext{e5}{e-mail: fabiocabral@uern.br}
\institute{Universidade de Bras\'ilia, Instituto de F\'isica, Caixa Postal 04455, Bras\'ilia, DF, 70919-970, Brazil \label{addr1}
\and 
Laboratoire d'Annecy de Physique des Particules (LAPP), Universit\'e Savoie Mont Blanc, CNRS/IN2P3, F-74941 Annecy, France \label{addr2} 
\and 
Centre for Cosmology, Particle Physics and Phenomenology, Institute of Mathematics and Physics, Louvain University, 2 Chemin du Cyclotron, 1348 Louvain-la-Neuve, Belgium \label{addr3}
\and 
Institut d'Astrophysique de Paris, GReCO, UMR7095 CNRS, 98 bis boulevard Arago, 75014 Paris, France \label{addr4}
\and
Universidade Estadual Paulista (UNESP), Instituto de Ci\^encia e Tecnologia, S\~ao Jos\'e dos Campos, SP, 12247-004, Brazil \label{addr5}
\and 
Instituto Nacional de Pesquisas Espaciais, Divis\~ao de Astrof\'\i sica, Av. dos Astronautas 1758, S\~ao Jos\'e dos Campos, SP, 12227-010, Brazil \label{addr6}
\and
Universidade do Estado do Rio Grande do Norte, Mossor\'o, RN, 59610-210, Brazil \label{addr7} 
}





\date{Received: date / Accepted: date}

\maketitle

\abstract{
The attempt to describe the recent accelerated expansion of the Universe includes different propositions for dark energy models and modified gravity theories. Establish their features in order to discriminate and even rule out part of these models using observational data is a fundamental issue of cosmology. In the present work we consider a class of extended theories of gravity (ETG) that are minimally coupled to the ordinary matter fields. In this context, and assuming a homogeneous and isotropic spacetime, we derive the energy conditions for this ETG class. We then put constraints on these conditions using a model-independent approach to reconstruct the deceleration function along with the Joint Light-curve Analysis (JLA) supernova sample, 11 baryon acoustic oscillation and 22 cosmic-chronometer measurements. We also consider an additional  condition imposing the strong energy condition  only on the ordinary matter. This is to guarantee the presence of only attractive matter in the energy-momentum tensor, at least in the redshift range of the observations, i.e., the recent accelerated expansion of the Universe is due solely to the modifications in the gravity theory. The main result of this work is a general reconstruction of the energy conditions valid for every considered ETG.
}
\section{Introduction}
\label{sec:introd}

In the last decades, a great amount of cosmological observational data
have been accumulated, endowing the modern cosmology with the capability
of quantitatively reproducing the details of many observed cosmic
phenomena, including the late time accelerating stage of the Universe.
The evidence of such an acceleration comes, for example, from the
measurements of the distance modulus of type Ia supernovae (SNe Ia),
however it still lacks a satisfactory physical explanation. Most of the current studies relates it to some unknown field (dark energy) or
to some extension of Einstein theory of General Relativity (GR).

In this context the energy conditions (ECs) are useful tools to evaluate some
features of the Universe evolution, since they can be derived considering
a minimum set of assumptions on the gravitation theory. The ECs have been
used to obtain more information from data assuming GR and homogeneous and isotropic metric
\citep{Schuecker2003, Santos2006, Gong2007, Santos2007, Lima2008,Lima2008a}. Among the interpretations of the ECs in GR, we have the
positivity of the energy density (weak EC) and the attractiveness of
gravity - focusing theorem - (strong EC) \cite{Wald1984}. In this sense, the accelerated expansion of the Universe, first evidenced by the SNe Ia observations~\cite{Riess1998,Perlmutter1999}, is an indication that the strong EC is currently being violated.

In refs.~\cite{Lima2008, Lima2008a} the authors suggested a methodology to analyze the fulfillment or not of the ECs in GR, by reconstructing the ECs from the 
SNe Ia data. As a result, they found a violation of the
strong EC with more than $99.73\%$ confidence interval in the redshift
range $[\sim 0.1, \, \sim 0.2]$. The ECs have also been addressed in the context of the Extended Theories of Gravity (ETGs) \cite{Capozziello2011,Nojiri2011,Capozziello2015}, such as $f(R)$ gravity, which modifies the Einstein-Hilbert Lagrangian by the introduction of an arbitrary function of the curvature scalar $R$. As shown by Santos et al. \cite{Santos2007b}, the ECs requirements may constrain the parameter space of each specific $f(R)$ functional form. On the other hand, Bertolami and Sequeira \cite{Bertolami2009} and Wang et al. \cite{Wang2010} have generalized the ECs for $f(R)$ theories with non-minimal coupling to matter. They found that the ECs are strongly dependent on the geometry. The conditions to keep the effective gravitational coupling positive as well as gravity attractive were also obtained. In a similar fashion, Wang and Liao \cite{Wang2012} have also obtained the ECs requirements for such a theory. Furthermore, imposing the fulfillment of the ECs at the present time they have constrained the parameter space for a particular Lagrangian using the current measurements of the Hubble parameter $H_0$, the deceleration parameter $q_0$ and the jerk parameter $j_0$. Recently, \citet{Santos2017} studied the attractive/non-attractive character of $f(R)$ gravity considering the strong energy condition.

As a consequence of the ECs derived for a generic $f(R)$ theory and of the equivalence between $f(R)$ gravity and scalar-tensor theories, Atazadeh et al. \cite{Atazadeh2009} have studied the ECs in Brans-Dicke theory and put constraints on the parameters involved. 

Using the parameters $H_0$, $q_0$ and $j_0$ together with the appropriate ECs, García et al. \cite{Garcia2011b}, Banijamali et al. \cite{Banijamali2012} and Atazadeh and Darabi \cite{Atazadeh2014} have shown the viability of some formulations of the modified Gauss-Bonnet gravity. The ECs inequalities were also used to constraint $f(R, T)$ theories \cite{Alvarenga2013}, where $T$ is the trace of the energy-momentum tensor. An extension of such a theory which also takes into account an arbitrary function of the quantity $Q = R_{ab}T^{ab}$ in the Lagrangian was considered in the scope of ECs by Sharif and Zubair \cite{Sharif2013b} and bounds on the parameters were obtained. Modified theories of gravity with non-null torsion have been also constrained with ECs. In this respect,  Sharif et al. \cite{Sharif2013} and Azizi and Gorjizadeh \cite{Azizi2017} were able to put bounds on the parameters of some particular formulations of the theory.

Besides, even the requirement that the ECs should be fulfilled or not has no clear meaning. 
As we will discuss later, the weak and dominant ECs are obtained by imposing direct restrictions on the energy-momentum tensor. In this work we assume that the matter-energy  content of the Universe is constituted just of attractive matter, i.e., the strong EC is violated only due to modifications in the gravity theory. Thus, to study the parameter space of the ETGs, it is reasonable to assume that both weak and dominant ECs are fulfilled throughout the cosmic history.  Differently, the null and strong ECs are derived from the Raychaudhuri equation, corresponding to impositions on the evolution of  null- and time-like congruences. At first, there is no observational evidence that leads us to assume that the null EC should not be fulfilled. On the other hand, the evidence of a recent accelerated expansion indicates that the strong EC has been violated. Since the null and strong ECs do not depend on the modified gravity terms (when assuming that the particles follow geodesics), in this work we will introduce a fifth condition to guarantee the presence of only attractive matter in the energy-momentum tensor.

In the majority of the work mentioned above, the constraints on the parameter space of each specific theory were obtained by imposing that the ECs are satisfied at the present time. However, such a statement does not imply the fulfillment (or not) of the ECs in the whole cosmic history. Consequently, the parameters could be further constrained if the ECs were extended for redshifts beyond $z = 0$. The aim of the present work is to give a general treatment for such an issue in the context of ETGs. 

In order to achieve this goal we consider a class of metric torsion-free
ETGs that presents a minimal curvature-matter coupling and the
energy-momentum tensor is conserved. We also assume a homogeneous and
isotropic spacetime and conformal transformation such that the extra
degrees of freedom of the theory can be written as an effective
energy-momentum tensor. This comprises the $f(R)$ theories, $f(R,T)$
theories for which the particular case $\nabla_b T^{ab} = 0$ is assumed, scalar-tensor
theories such as Brans-Dicke theory and several other possible
formulations \cite{Capozziello2011}. We write the ECs and the fifth condition in terms of the modified gravity functions. Then we apply the model-independent reconstruction method of the
deceleration function introduced by \citet{Vitenti2015} to obtain
observational constraints on the functions of the modified term, given
the EC and the additional inequations, for a redshift range and not only
for the present time. In particular, we use the Joint Light-curve
Analysis (JLA) supernova sample \cite{Betoule2014}, 11 baryon acoustic
oscillation (BAO) data points from 6dF Galaxy Survey and the Sloan
Digital Sky Survey (SDSS)~\cite{Beutler2011, Ross2015, Alam2017, Bautista2017, Ata2018} and 22 cosmic-chronometer [$H(z)$]
measurements~\cite{Stern2010, Moresco2012, Moresco2015, Riess2016}.
 
The layout of the paper is as follows. In Sect.~\ref{sec:EC} we introduce
the ECs in the context of a class of ETGs, and then we derive the EC
inequations considering a homogeneous and isotropic spacetime. In
Sect.~\ref{sec:recons_qz} we recall the main steps of the
model-independent reconstruction approach \cite{Vitenti2015}, and obtain
the deceleration and Hubble function estimates using SNe Ia, BAO and
$H(z)$ data. The observational bounds on the ECs and the additional
condition and their respective implications in the context of GR and on
the ETG functions are discussed in Sect.~\ref{sec:results}. Finally, we
present our conclusions in Sect.~\ref{conclusions}. Throughout the article
we adopt the metric signature $(-,+,+,+)$.

\section{Energy Conditions}
\label{sec:EC}

In this section, we define the ECs in the context of a class of ETGs for which the field equations can be written in the following generic form \citep{Capozziello2015} 
\begin{equation}\label{eq:field_eqs}
g_1(\Psi^i) \left(G_{ab} + H_{ab} \right) = 8\pi G g_2(\Psi^i) T_{ab},
\end{equation}
where $G_{ab} \equiv R_{ab} - \frac{1}{2}g_{ab} R$, $R_{ab}$ is the Ricci tensor, $R = R^a_{\ a}$ is the Ricci scalar, $G$ is the gravitational constant, and $T_{ab}$ is the energy-momentum tensor of the matter fields. The tensor $H_{ab}$ encapsulates the additional geometrical information of the modified theory. For instance, it may depend on scalar and (or) vector fields, scalars made out of Riemann and Ricci tensors, and derivatives of these quantities (see \cite{Felice2010,Capozziello2011} and references therein). Finally, $\Psi^i$ refers to these mentioned fields and geometric quantities, and the modified coupling with the matter fields is given by $g_1(\Psi^i)$ and $g_2(\Psi^i)$, where the latter includes explicit curvature-matter couplings \cite{Harko2014,Capozziello2015}. 

In this work we consider a class of ETGs that presents a minimal curvature-matter coupling\footnote{See \cite{Harko2014,Capozziello2015} for a discussion of nonminimal curvature-matter coupling.}, i.e., $g_2(\Psi^i) = 1$, and the matter action is invariant under diffeomorphisms, then $\nabla^a T_{ab} = 0$ \cite{Wald1984}. Given the conservation of the energy-momentum tensor and the twice-contracted Bianchi identity, $\nabla^a G_{ab} = 0$, we have that \cite{Capozziello2015}
\begin{equation}
\nabla^a H_{ab} = - \frac{8\pi G}{g_1^2} T_{ab} \nabla^{a} g_1.
\end{equation} 

Many authors have been discussing the ECs in the context of different ETGs such as $f(R)$ \cite{Santos2007b,Bertolami2009,Wang2010,Wang2012,Santos2017}, scalar-tensor gravity theories \cite{Sharif2013} and massive gravity \cite{Alves2017}. A common procedure, though, is to consider the modified gravity term
$H_{ab}$ as a source of the  {\it effective} energy-momentum tensor,
\begin{equation}\label{eq:Teff}
T_{ab}^{\rm eff} = \frac{T_{ab}}{g_1} - \frac{H_{ab}}{8\pi G}.
\end{equation}

However, as also pointed out by \citet{Capozziello2015}, these fictitious fluids can be related to, for example, scalars constructed from geometrical quantities or other further degrees of freedom. In this case, it could be somewhat misleading to apply the standard ECs obtained from GR to the resulting effective energy-momentum tensors derived in such theories. Hence, one should perform suitable conformal transformations in
order to better define the energy conditions in terms of $T_{ab}^{\rm
eff}$ \cite{Carloni2009, Capozziello2014, Capozziello2015}. Moreover, as we will see bellow, two energy conditions (strong and null) are related to the convergence conditions and have specific interpretations in a GR setting~\cite{Hawking1973}. Such interpretations are, in general, distinct in the context of a ETG.

Notwithstanding, in the present article we avoid these misleading issues by using the convergence conditions directly in the definitions of the strong and null ECs. This is because our main concern is about which bounds can be imposed in an alternative theory of gravity in the hypothesis that the accelerated expansion of the Universe is uniquely due to its extra terms that modify the GR theory. In other words, if a given ETG is able to explain the accelerated expansion, then there is no need to include any extra fluids in the cosmological model. As it will be clear in what follows, this is equivalent to say that the ECs should be fulfilled by the energy-momentum tensor of the current matter content of the Universe, at least in the redshift range covered by the cosmological observations. Obviously, there is no guarantee that anyone of these conditions could be violated in other cosmological epochs as, for example, in the early inflationary period.

Therefore, from the above argumentation we make clear what we mean by ``ECs bounds on ETG'': they are bounds that an ETG needs to respect in order to avoid the inclusion of a dark fluid to accelerate the expansion of the Universe.

\subsection{Energy conditions for ETGs with minimal coupling}\label{sec:ecetg}

Let us start with the above mentioned ECs and define them from the expansion rate of the Universe. Hence, consider a timelike geodesic congruence with a tangent vector field $t^a$ and let $\tau$ be a parameter of these timelike curves. As matter is minimally coupled to geometry, test particles follow geodesics, i.e., we are assuming that the modified gravity does not introduce any additional force. In this case the Raychaudhuri equation is given by
\begin{equation}\label{eq:raychaudhuri_1}
\frac{d\theta}{d\tau} = -\frac{1}{3}\theta^2 - \sigma_{ab}\sigma^{ab}
+ \omega_{ab}\omega^{ab} - R_{ab}t^a t^b ,
\end{equation}
where $\theta = \nabla_a t^a$ is the expansion of the congruence. Analogously, for a congruence of null curves parametrized by $\lambda$ and with tangent vector $k^a$, we have
\begin{equation}\label{eq:raychaudhuri_2}
\frac{d \hat{\theta}}{d\lambda} = -\frac{1}{2} \hat{\theta}^2 - \hat{\sigma}_{ab} \hat{\sigma}^{ab} + \hat{\omega}_{ab}\hat{\omega}^{ab} - R_{ab} k^{a}k^{b}.
\end{equation}

In the above equations, $\sigma_{ab}$ ($\hat{\sigma}_{ab}$) is the shear tensor, $\omega_{ab}$ ($\hat{\omega}_{ab}$) is the vorticity tensor, and the hat means that these quantities are projected onto the subspace normal to the null vectors. Note that the necessary and sufficient conditions for a congruence be hypersurface-orthogonal is $\omega_{ab} = \hat{\omega}_{ab} = 0$ \cite{Wald1984}. Considering null vorticity and given that the second term of Eq.~\eqref{eq:raychaudhuri_1} is nonpositive, the convergence of timelike geodesics (i.e., focusing) occurs if
\begin{equation}\label{eq:SEC_R}
R_{ab}t^a t^b \geq 0,
\end{equation}
this inequation is also known as \emph{timelike convergence
condition}~\cite{Hawking1973}.

Thus, by using Eq.~\eqref{eq:field_eqs} with $g_2 = 1$, we find what we call here the {\it strong energy condition} (SEC) in an ETG
\begin{equation}\label{eq:SEC_ETG}
{\rm{\bf{SEC:}}} ~~~\left[\frac{8\pi G}{g_1} \left(T_{ab}t^{a}t^{b} - \frac{T}{2}\right) - \left(H_{ab}t^{a}t^{b} - \frac{H}{2}\right) \right]  \geq 0,
\end{equation}
which expresses the attractive character of gravity. In this case, as pointed out by Capozziello {\it et al.} \cite{Capozziello2015}, even if the matter fields do not contribute positively (e.g., a matter field with negative pressure), Eq.~\eqref{eq:SEC_ETG} can still be fulfilled given the geometric term. In other words, the attractiveness of gravity can remain in the presence of a {\it dark energy} (DE) like fluid depending on the modified gravity theory. Of course, this is not the case as indicated by the cosmological observations, and the above defined SEC is violated, for this reason this bound serves to measure with what statistical confidence the violation occurs.

Similarly, from Eq.~\eqref{eq:raychaudhuri_2} we have that the condition for the convergence of null geodesics is $R_{ab}k^a k^b \geq 0$, which is the \emph{null convergence condition}~\cite{Hawking1973}. Thus, the {\it null energy condition} (NEC) in the context of an ETG is given by the following inequality
\begin{equation}\label{eq:NEC_ETG}
{\rm{\bf{NEC:}}} ~~~\frac{8 \pi G}{g_1} T_{ab}k^a k^b - H_{ab}k^a k^b \geq 0.
\end{equation}

Therefore, notice that in the way the SEC and the NEC are defined, they are not conditions only on the energy-momentum tensor but in a sum of the energy-momentum tensor with extra terms of the modified gravity, a quite different situation to what happens in the GR theory. In the GR case we have $g_1 = 1$ and $H_{ab} = H = 0$, and the above conditions are written just in terms of $T_{ab}$.

In contrast, the {\it weak} and {\it dominant energy conditions}, WEC and DEC, respectively, are {\it direct} restrictions on the energy-momentum tensor, $T_{ab}$, for any theory of gravity (they do not originate from the Raychaudhuri equation). They are quite reasonable physical conditions expected to be satisfied for the mean energy-momentum tensor of the matter that fills the Universe. As we will see in more details in Sect.~\ref{sec:HI}, the WEC states that the matter energy density is positive for every time-like vector, i.e.,
\begin{equation}\label{eq:WEC_T}
{\rm \bf{WEC:}}~~~T_{ab}t^a t^b \geq 0,
\end{equation}
and the DEC states that the speed of the energy flow of matter is less than the speed of light. This condition can be written in the form
\begin{equation}\label{eq:DEC_T}
{\rm \bf{DEC:}}~~~T_{ac}{T^c}_b t^a t^b \leq 0.
\end{equation}

In the present work we compare the above bounds with the
reconstruction of the geometry. This means that we estimate
$R_{ab}$ directly from the data and then apply the bounds to it. For this
reason, the ECs stemming directly from $T_{ab}$ (WEC and
DEC) will depend explicitly on the ETG functions, while SEC and NEC will not depend on them. 

Now, let us introduce a fifth energy condition (FEC) on the energy-momentum as follows
\begin{equation}\label{eq:SEC:GR}
{\rm \bf{FEC:}}~~~\frac{8\pi G}{g_1} \left( T_{ab} t^a t^b - \frac{1}{2} T \right) \geq 0.
\end{equation}

This is just the SEC in the GR theory (taking $g_1 = 1$). On the other hand, in the context of an ETG the meaning of this condition is that if the FEC is fulfilled, than the violation of the SEC (\ref{eq:SEC_ETG}) is due solely to the term $H_{ab}$, i.e., this is the only term responsible for the acceleration of the Universe. Of course, it is well understood that there is no prior reason for a given energy-momentum tensor fulfill the condition (\ref{eq:SEC:GR}). A classical example is the energy-momentum tensor of a scalar field minimally coupled to gravity, for which this condition can be violated even for the case of a massive potential. However, if a certain ETG intends to solve the DE problem without the introduction of any additional fluid, then the FEC needs to be imposed on the energy-momentum tensor describing the matter content of the Universe, at least in the redshift range covered by the observations. If the FEC is not fulfilled, than we can say that the theory still requires a negative pressure fluid to explain the accelerated expansion.

\subsection{Homogeneous and isotropic spacetime}
\label{sec:HI}

In this section we derive the ECs for a homogeneous and isotropic Universe. This is described by the Friedmann-Lema\^{\i}tre-Robertson-Walker (FLRW) metric,
\begin{equation}
\label{eq:rw_metric} 
ds^2 = -c^2\,dt^2 + a^2 (t) \left [ dr^2 + S_k^2(r) (d\theta^2 + \sin^2 \theta  d\phi^2) \right ]\;,
\end{equation}
where $a(t)$ is the scale factor and $k=0, 1$ or $-1$, whose flat, spherical and hyperbolic functions are $S_k(r)=(r\,$, $\sin(r)$, $\sinh(r))$, respectively. In this case, $R_{ab}$ and $H_{ab}$ are diagonal tensors and their components are functions of the time $t$ only, namely, 
\begin{align}
R_{00} &= -3 \frac{\ddot{a}}{a}, & R_{ij} &= \left[ \frac{\ddot{a}}{a} + 2\left(\frac{\dot{a}}{a} \right)^2 + k \right]g_{ij},
\end{align}
and, analogously,
\begin{equation}\label{eq:Hab}
H_{ab} = (\htt(t) g_{00}, \hs(t) g_{ij}).
\end{equation}
As stated above, to be compatible with a Friedmann metric, the tensor $H_{ab}$ must have the form of Eq.~\eqref{eq:Hab}. This means that all information about the ETG is encoded in two time dependent functions $\htt(t)$ and $\hs(t)$.

In turn, the energy-momentum tensor for the matter fields can be written as 
\begin{equation}\label{eq:em_tensor}
T_{ab} = (\rho + p)U_a U_b + pg_{ab},
\end{equation}
where $\rho$ is the matter-energy density, $p$ is the pressure, and the
four-velocity of the fluid is $U^a$ (where $U_aU^a=-1$). Hence, the
Friedmann equations for the considered ETGs acquire the following form,
\begin{align}
\left( \frac{\dot{a}}{a} \right)^2 + \frac{k}{a^2} - \frac{1}{3} \htt &= \frac{8\pi G}{3g_1} \rho. \label{eq:Friedmann2}\\
-3\frac{\ddot{a}}{a} - \frac{1}{2} (\htt - 3\hs) &= \frac{4\pi G}{g_1} (\rho + 3p), \label{eq:Friedmann1}
\end{align}

Finally, giving a normalized timelike vector $t^a = NU^a + N^a$, where $t_at^a = -1$ and $U_aN^a = 0$, and a null vector $k^a$, we rewrite the ECs by substituting Eqs.~\eqref{eq:Hab} and \eqref{eq:em_tensor} into Eqs.~\eqref{eq:SEC_ETG}, \eqref{eq:NEC_ETG} \eqref{eq:WEC_T} and \eqref{eq:DEC_T}, thus
\begin{align}
&{\rm {\bf NEC}} \quad \frac{8\pi G}{g_1}(\rho + p) + \htt - \hs \geq 0, \label{eq:NEC_a} \\
 \nonumber \\
&{\rm {\bf SEC}} \quad \frac{8\pi G}{g_1}(\rho + 3p) + \htt - 3\hs \geq 0  \quad {\rm and} \label{eq:SEC_a} \\
& \quad \qquad \frac{8\pi G}{g_1}(\rho + p) + \htt - \hs \geq 0,\nonumber \\
\nonumber\\
&{\rm {\bf WEC}} \quad \rho \geq 0 \quad {\rm and} \quad \rho + p \geq 0, \label{eq:WEC_a}\\
\nonumber\\
&{\rm {\bf DEC}} \quad -\rho \leq p \leq \rho \quad {\rm and} \quad \rho \geq 0  \label{eq:DEC_a}. 
\end{align}

Note that, imposing that the conditions are fulfilled for
any timelike/null vector field (parameterized as the timelike vector
field $t_a$ defined above), we get two conditions for
SEC, WEC and DEC when we apply it to a energy-momentum tensor compatible
with a Friedmann metric.

In order to confront the ECs with observational data, such that one can infer local (in redshift) information about the fulfillment of these conditions, Lima et al.~\cite{Lima2008,Lima2008a} showed that it is convenient to write the ECs in terms of the Hubble function, $H(z) \equiv \dot{a}/a = H_0 E(z)$, and the deceleration function, $q(z) \equiv -\ddot{a}a/{\dot{a}^2}$, where $1+z = a_0/a$. Therefore, by using the Friedmann equations, Eqs.~\eqref{eq:Friedmann1} and \eqref{eq:Friedmann2}, the energy conditions are rewritten as
\begin{align}
& {\rm {\bf NEC}} \quad \left[1 + q(z)\right] E(z)^2 - \Omega_k^0 (1+z)^2 \geq 0, \label{eq:NEC}\\
\nonumber\\
& {\rm {\bf SEC}} \quad q(z) \geq 0, \label{eq:SEC}\\
\nonumber \\
& {\rm {\bf WEC 1}}\quad \frac{\htt}{3H_0^2} \leq E(z)^2 - \Omega_k^0 (1+z)^2, \label{eq:WEC1}\\
\nonumber\\
& {\rm {\bf WEC 2}} \quad \frac{(\htt - \hs)}{2H_0^2} \leq [1 + q(z)]E(z)^2 - \Omega_k^0(1+z)^2, \label{eq:WEC2}\\
\nonumber\\
& {\rm {\bf DEC}} \quad \frac{(\htt + \hs)}{6H_0^2} \leq \frac{[2 - q(z)]E(z)^2 - 2\Omega_k^0(1+z)^2}{3}, \label{eq:DEC}
\end{align}
where $\Omega_k^0 = -k/(a_0H_0)^2$ and the subscript (superscript) 0 stands for the present-day quantities. Conditions WEC1 and WEC2 refer respectively to the first and second inequalities in~\eqref{eq:WEC_a}. Except for WEC, the above expressions refer only to the first inequation of each condition in Eqs.~(\ref{eq:SEC_a}--\ref{eq:DEC_a}) ($\rho - p \geq 0$ for DEC), and provide a complete unambiguous set of conditions.

The recent accelerated expansion of the Universe evinced by SNe Ia~\cite{Riess1998,Perlmutter1999}, large scale structure (LSS) \cite{Tegmark2006,DES2017} and cosmic microwave background radiation (CMB) \cite{Hinshaw2013, Planck2015} data represents the violation of SEC \cite{Lima2008, Lima2008a}, i.e., $q(z) < 0$. As a consequence, this fact requires a modified theory of gravity and/or the existence of an exotic fluid.

In this sense, we have defined the FEC in the end of Sect.~\ref{sec:ecetg} which implies that the Universe if filled only of ordinary attractive matter. Thereby, from Eq.~(\ref{eq:SEC:GR}) we state that
\begin{equation}
{\rm {\bf FEC}}~~~~~\frac{8\pi G}{g_1}(\rho + 3p) \geq 0.
\end{equation}

That is, any contribution for a negative value of the deceleration
function, and consequently a DE like behavior, originates
exclusively from the modified gravitational term $H_{ab}$, thus
\begin{align}
\label{eq:fifth_cond}
{\rm {\bf FEC}}~~~~~\frac{\left(-\htt + 3\hs\right)}{6H_0^2} &\geq  -q(z)E(z)^2.
\end{align}

One could also think that a similar new condition could be
obtained from NEC. However, that is not the case, such a condition
coincides exactly with the second one obtained from WEC, i.e., WEC2. It is easy to observe this by comparing the second condition in Eq.~\eqref{eq:WEC_a} with Eq.~\eqref{eq:NEC_a}.

In this work we aim to put observational constraints on combinations of $\htt(z)$ and $\hs(z)$ by requiring the fulfillment of the WEC, DEC and FEC. For this, it is clear that Eqs.~(\ref{eq:NEC}--\ref{eq:fifth_cond}) require estimates of $q(z)$ and $E(z)$. Hence in the following section we will describe the methodology and observational data sets to obtain these estimates. 

\begin{table}[h]
\caption{BAO data} 
\label{tab:bao}
\centering
\begin{threeparttable}[b]
{\begin{tabular}{l c c c c rrrrrr} 
\hline 
Measurement & redshift & mean & $\sigma$ & \multicolumn{6}{c}{$\mathsf{C}_{\text{BAO}}$}  \\[0.5ex]
\hline
$r_{\rm d}/D_{\rm V}(z) \,$\tnote{a} & 0.106 & 0.336 & 0.015 & &  &  &  &  &   \\ 
\hline
$D_{\rm V}(z)/r_{\rm d} \,$\tnote{b} & 0.15 & 4.466 & 0.168 & & &  & & & & 
\\
\hline
$D_{\rm M}(z) \,$\tnote{c} & \multirow{2}{*}{0.38} & 1518 & 22 & 484.0 & 9.53 & 295.2 & 4.67 & 140.2 & 2.40 
\\
$H(z)$ & & 81.5 & 1.9 & 9.53 & 3.61 & 7.88 & 1.76 & 5.98 & 0.92 & \\
$D_{\rm M}(z)$ & \multirow{2}{*}{0.51} & 1977 & 27 & 295.2 & 7.88 & 729.0 & 11.93 & 442.4 & 6.87 & \\
$H(z)$ & & 90.4 & 1.9 & 4.67 & 1.76 & 11.93 & 3.61 & 9.55 & 2.17 & \\
$D_{\rm M}(z)$ & \multirow{2}{*}{0.61} & 2283 & 32 & 140.2 & 5.98 & 442.4 & 9.55 & 1024.0 & 16.18 & \\
$H(z)$ & & 97.3 & 2.1 & 2.40 & 0.92 & 6.87 & 2.17 & 16.18 & 4.41 & \\
\hline
$D_{\rm V}(z)/r_{\rm d} \,$\tnote{d} & 1.52 & 26.086 & 1.15 & & & &  & & 
\\
\hline
$D_{\rm H}(z)/r_{\rm d} \,$\tnote{e} & \multirow{2}{*}{2.33} & 9.07 & 0.31 & & & & & & 
\\
$D_{\rm M}(z)/r_{\rm d}$ & & 37.77 & 2.13 & & & & & & & 
\\[1ex] 
\hline
\end{tabular}}
\begin{tablenotes}
	\item[a]{Beutler et al.~\cite{Beutler2011}.}
    \item[b] \citet{Ross2015} provides $D_V(z)(r_d^{\mathrm{fid}}/r_d)$, where $r_d^{\mathrm{fid}} = 148.69 \, h^{-1}\text{Mpc}$. We consider this fiducial value to build $L_{\text{BAO}}$.
    \item[c] Alam et al.~\cite{Alam2017}.
    \item[d] Ata et al.~\cite{Ata2018}.
    \item[e] Analogously, \citet{Bautista2017} work with the variables $\alpha_\parallel = \frac{\left[D_{\rm H}(z)/r_{\rm d} \right]}{\left[D_{\rm H}(z)/r_{\rm d} \right]_{\mathrm{fid}}}$ and $\alpha_\perp = \frac{\left[D_{\rm M}(z)/r_{\rm d} \right]}{\left[D_{\rm M}(z)/r_{\rm d} \right]_{\mathrm{fid}}}$, where $\left[D_{\rm H}(z)/r_{\rm d} \right]_{\mathrm{fid}} = 8.612$ and $\left[D_{\rm M}(z)/r_{\rm d} \right]_{\mathrm{fid}} = 39.15$.
  \end{tablenotes}
\end{threeparttable}
\end{table}

\section{Model-independent estimates of $q(z)$ and $E(z)$}
\label{sec:recons_qz}

Here we apply the model-independent method to reconstruct the deceleration function introduced by Vitenti and Penna-Lima \cite{Vitenti2015} (hereafter VPL), using SNe Ia, BAO and H(z) data (see section~\ref{sec:data}). The approach is model-independent because $q(z)$ is reconstructed without specifying the matter content of the Universe or the theory of gravitation. Particularly, we just assume that the Universe is homogeneous and isotropic.

In short, the VPL approach consists in approximating the $q(z)$ function by a cubic spline over the redshift range $[z_{\text{min}}, z_{\text{max}}]$, where the minimum and maximum redshifts are defined by the observational data. Then, choosing the number of knots, $n+1$, we write the reconstructed curve $\hat{q}(z)$ in terms of the parameters $\{\hat{q}_i\}$, $i = 0, ..., n$. As discussed in Ref.~\cite{Vitenti2015}, the complexity of the reconstructed function depends on $n$. However, instead of varying the number of knots, VPL addressed this question by including a penalty function, which is parametrized by $\sigma_{\rm rel}$, such that small values of $\sigma_{\rm rel}$ (e.g., $\sigma_{\rm rel} = 0.05$) force $\hat{q}(z)$ to be a linear function, whereas large values (e.g., $\sigma_{\rm rel} = 1.5$) provide a high-complexity function. By construction, the errors of the reconstructed curves are dominated by biases (small $\sigma_{\rm rel}$) and over-fitting (large $\sigma_{\rm rel}$).

Considering different values of $\sigma_{\rm rel}$ and fiducial models for $q(z)$, VPL validated the reconstruction method via the Monte Carlo approach, using SNe Ia, BAO and $H(z)$ mock catalogs. Then, 
evaluating the bias-variance trade-off for each case, which is characterized by a fiducial model and a $\sigma_{\rm rel}$ value, the best reconstruction method was determined. That is, the best $\sigma_{\rm rel}$ value is the one that minimizes the mean squared error, requiring the bias to be at most 10\% of this error.

Finally, here we use the VPL approach considering 12 knots, $\sigma_{\rm rel} = 0.3$ and the redshift interval $[0.0, 2.33]$ to reconstruct the deceleration function (for further discussions, see VPL~\cite{Vitenti2015}). The observable quantities such as $E(z)$ and the transverse comoving distance $D_M(z)$ are written, respectively, in terms of $q(z)$ as
\begin{equation}
E(z) = \frac{H(z)}{H_0} = \exp \int_0^z \frac{1+ q(z^\prime)}{1+z^\prime} dz^\prime,
\end{equation}
and
\begin{equation}
D_M(z) = \left\{
    \begin{array}{ll}
        K^{-1} \sin\left(K \chi (z)\right)   &\mbox{for $\Omega_k < 0$,} \\
        \chi (z)                             &\mbox{for $\Omega_k = 0$,} \\
        K^{-1} \sinh\left(K \chi (z)\right)  &\mbox{for $\Omega_k > 0$,}
    \end{array}
\right.
\end{equation}
where $K = \frac{H_0 \sqrt{\vert\Omega^0_k\vert}}{c}$ and the comoving distance is
\begin{equation}
\chi (z) = \frac{c}{H_0} \int_0^z \frac{dz^\prime}{E(z^\prime)}. 
\end{equation}
 
\subsection{Data}
\label{sec:data}

In the present work, we use some current available observational data for small redshifts ($z\leq 2.33$) and their likelihoods, namely, the Sloan Digital Sky Survey-II and Supernova Legacy Survey 3 years (SDSS-II/SNLS3) combined  with Joint Light-curve Analysis (JLA) SNe Ia sample \cite{Betoule2014}, BAO data \cite{Beutler2011, Ross2015, Alam2017, Bautista2017, Ata2018} and $H(z)$ measurements \cite{Riess2016, Moresco2012, Moresco2015, Stern2010}:
\begin{equation}\label{eq:full_likelihood}
-2\ln L(\vec{d}, \boldsymbol{\theta}) = -2\left(\ln L_{\rm SNIa} + \ln L_{\rm BAO} + \ln L_{\rm H}\right) + \sum_{i = 1}^{10} \left( \frac{\overline{\hat{q}}_i - q_i}{0.3 + 10^{-5} \vert\overline{\hat{q}}_i\vert} \right)^2,
\end{equation}
where $\vec{d}$ and $\boldsymbol{\theta}$ comprehend the observational data sets and the parameters to be fitted, respectively, and the last term is the penalization factor with $\overline{\hat{q}}_i = (\hat{q}_{i-1} + \hat{q}_{i+1})/2$.

The JLA SNe Ia likelihood is given by 
\begin{equation}
-2\ln L_{\rm SNIa} =  (\vec{m}_{\rm B} - \vec{m}_{\rm B}^{\rm th})^T C_{\rm SNIa}^{-1} (\vec{m}_{\rm B} - \vec{m}_{\rm B}^{\rm th}) + \ln \vert C_{\rm SNIa} \vert,
\end{equation}
where $\vec{m}_{\rm B}$ is a vector of the 740 measured rest-frame peak B-band magnitudes, $C_{\rm SNIa}^{-1}$ and $\vert C_{\rm SNIa} \vert$ are the inverse and the determinant of the covariance matrix, respectively, and 
\begin{equation}
m_{\rm Bi}^{\rm th} = 5\log_{10} D_L(z^{\rm hel}_i, z^{\rm cmb}_i) - \alpha X_i + \beta C_i + M_{h_i} - 5\log_{10}(c/H_0) + 25.
\end{equation}
The SNe Ia astrophysical parameters $\alpha, \beta, M_1$ and $M_2$ are related to the stretch-luminosity, colour-luminosity and the absolute magnitudes, respectively. The luminosity distance is $D_L(z^{\rm hel}, z^{\rm cmb}) = (1 + z^{\rm hel}) D_M(z^{\rm cmb})$, where $z^{\rm hel}$ and $z^{\rm cmb}$ are the heliocentric and CMB frame redshits.
 
The BAO likelihood is
\begin{equation}\label{eq:bao}
-2 \ln L_{\rm BAO} = \left[\vec{b}^{\rm th}(z) - \vec{b}\right]^T C_{BAO}^{-1} 
\left[\vec{b}^{\rm th}(z) - \vec{b}\right] 
- 2\ln L_{\mathrm{Ross}} - 2\ln L_{\mathrm{Bautista}},
\end{equation} 
where $\vec{b}$ and $C_{BAO}$ represent, respectively, 8 BAO measurements and the respective covariance matrix \cite{Beutler2011, Alam2017,Ata2018} as displayed in Table~\ref{tab:bao}. The $\vec{b}^{\rm th}(z)$ vector is composed of the quantities (and their combinations) $D_M(z)$, $H(z)$, $D_H (z) = c/H(z)$,
\begin{equation}
D_V(z) \equiv \left[ z D_M(z)^2 D_H(z) \right]^{1/3},
\end{equation}
and $r_d$, i.e., the sound horizon at the drag redshift,
\begin{equation}
r_d \equiv r_s (z_d) = \frac{1}{H_0} \int_{z_d}^{\infty} dz \frac{c_s(z)}{E(z)},
\end{equation}
where $c_s(z)$ is the sound wave speed in the photon-baryon fluid.
Since the VPL reconstruction method is defined in a small
redshift interval, we cannot calculate the integral above. Furthermore,
this integral is model dependent, and for this reason in this analysis we
treat $r_d$ as a free parameter. The last two terms of
Eq.~\eqref{eq:bao} are computed using the likelihood distribution from
Refs.~\cite{Ross2015,Bautista2017}, respectively.
 
We use 22 $H(z)$ measurements obtained from Refs.~\cite{Riess2016, Moresco2012, Moresco2015, Stern2010} through the cosmic chronometers formalism. The $H(z)$ likelihood is
\begin{equation}
-2 \ln L_{\rm H} = \sum_{i = 1}^{22} \frac{\left( H(z_i) - H^{\rm obs}_i\right)}{\sigma_i^2},
\end{equation}
where the data points $H^{\rm obs}_i$, errors $\sigma_i$ and the respective references are listed in Table~\ref{tab:Hz}.

\begin{table}[h]
\caption{Cosmic-chronometer data} 
\label{tab:Hz}
\centering
\begin{threeparttable}[b]
\begin{tabular}{c c c c | c c c c}
	\hline
redshift & $H(z)$ & $\sigma$ & reference & redshift & $H(z)$ & $\sigma$ & reference \\[0.5ex]
\hline
0.0 & 73 & 1.75 & Riess et al.~\cite{Riess2016} 
& 0.1 & 69 & 12 & \multirow{11}{*}{Stern et al.~\cite{Stern2010}} \\  
0.18 & 75 & 4 & \multirow{8}{*}{Moresco et al.~\cite{Moresco2012}} & 0.17 & 83 & 8 & \\
0.20 & 75 & 5 & & 0.27 & 77 & 14 & \\
0.35 & 83 & 14 & & 0.4 & 95 & 17 & \\
0.59 & 104 & 13 & & 0.48 & 97 & 60 & \\
0.68 & 92 & 8 & & 0.88 & 90 & 40 & \\
0.78 & 105 & 12 & & 0.9 & 117 & 23 & \\
0.88 & 125 & 17 & & 1.3 & 168 & 17 & \\
1.04 & 154 & 20 & & 1.43 & 177 & 18 & \\
1.363 & 160 & 33.6 & \multirow{2}{*}{Moresco et al.~\cite{Moresco2015}} & 1.53 & 140 & 14 &\\
1.965 & 186.5 & 50.4 & & 1.75 & 202 & 40 &\\[1ex]
\hline
\end{tabular}
\end{threeparttable}
\end{table}

\subsection{Analysis}
\label{sec:analysis}

We now apply the VPL methodology to reconstruct $q(z)$ (and, consequently, $H(z)$ and the cosmological distances) along with the observational data, described in Sects.~\ref{sec:recons_qz} and \ref{sec:data}, respectively, to put constraints on $\htt(z)$ and $\hs(z)$ from the EC bounds (see Eqs.~\eqref{eq:WEC1}, \eqref{eq:WEC2}, \eqref{eq:DEC} and \eqref{eq:fifth_cond}). We also evaluate the violation/fulfillment of the ECs in GR. The numerical as well as the post-processing analyses carried out in this work made use of the Numerical Cosmology library ({\sf NumCosmo})~\cite{DiasPintoVitenti2014}. 

First, following the same procedure as in VPL, we use the {\sf NcmFitESMCMC} function to perform a Markov Chain Monte Carlo (MCMC) analysis given an ensemble sampler with affine invariance approach \cite{Goodman2010}. Thus, from Eq.~\eqref{eq:full_likelihood} and to avoid any further assumptions on the astrophysical and cosmological dependencies of the SNe Ia, BAO and $H(z)$ likelihoods, we reconstruct $q(z)$ by fitting  its coefficients along with $\Omega_k^0$, the SNe Ia parameters, the drag scale (present in the BAO likelihood) and the Hubble parameter $H_0$, i.e.,
\begin{equation}
\theta = \{\{\hat{q}_i\}, \Omega_k^0, \alpha, \beta, M_1, M_2, H_0, r_d \},
\end{equation}
where $i = 0, ..., 11$.

Here we consider three different cases regarding the spatial curvature. In the first two we fit $\Omega_k^0$ by assuming zero-mean Gaussian priors with standard deviation equal to $0.05$ and $0.10$, respectively. This choice is consistent with the {\it Planck} results \cite{Planck2015}. The third case refers to a flat Universe, i.e., we fix $\Omega_k^0 = 0.0$. As described in Sect.~\ref{sec:recons_qz}, the ``prior'' on $\{\hat{q}_i\}$ corresponds to the penalization factor in Eq.~\eqref{eq:full_likelihood}. By construction, the Riess et al.~\cite{Riess2016} data point is a prior for $H_0$. Finally, we assume flat priors for $\{\alpha, \beta, M_1, M_2\}$, given the respective ranges: $\alpha$ and $\beta \in [0, \, 5]$, $M_1$ and $M_2 \in [-30, \, -10]$. 

Thus, for each case, we ran the {\sf NcmFitESMCMC} algorithm computing $5
\times 10^6$ sampling points in the 18- and 19-dimensional  parameter
spaces distributed among 50 chains. The convergence is attained as
indicated by the multivariate potential scale reduction factor (MPSRF) of
about 1.015 and the effective sample size (see \cite{Doux2017} for
details of the convergence tests and criteria). The variance of $-2\ln L$
and all 19 parameters (18 in the flat case) also converged, e.g.,
$\mathrm{Var}(-2\ln L) \simeq 36.08$ (flat case) which is consistent
with a chi-squared distribution with 18 degrees of freedom
($\chi_{18}^2$).

\section{Results}
\label{sec:results}

\begin{figure}
	\centering
\includegraphics[scale=0.90]{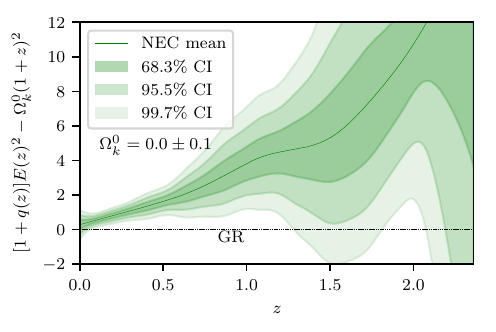}
\includegraphics[scale=0.90]{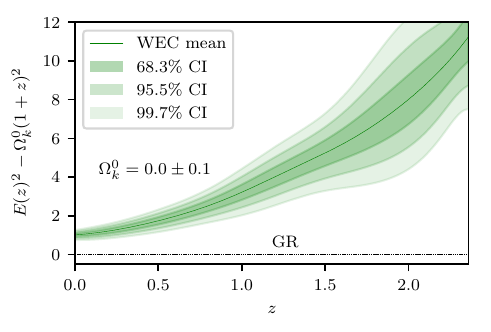}%
\includegraphics[scale=0.90]{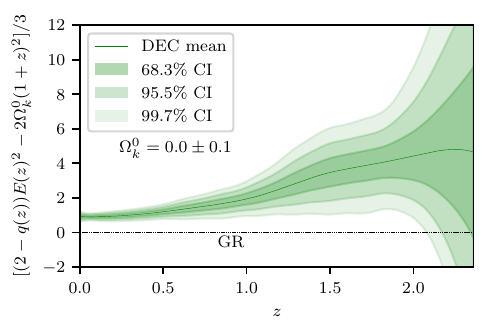}
\caption{\label{fig:ec_gr} The green solid lines are the mean estimates of the NEC (upper panel), WEC (left-lower panel) and DEC (right-lower panel). The green shaded areas are their 68.3\%, 95.5\% and 99.7\% CIs at different redshift values obtained from the reconstructed $q(z)$ by using SNe Ia + BAO + $H(z)$ data. These MCMC results were obtained using a Gaussian prior on $\Omega_k^0$ with zero mean and $\sigma = 0.1$. The black dashed line indicates the GR bounds.}
\end{figure}

\begin{figure}
\includegraphics[scale=0.89]{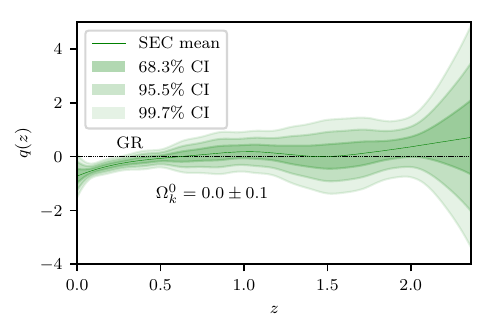}%
\includegraphics[scale=0.89]{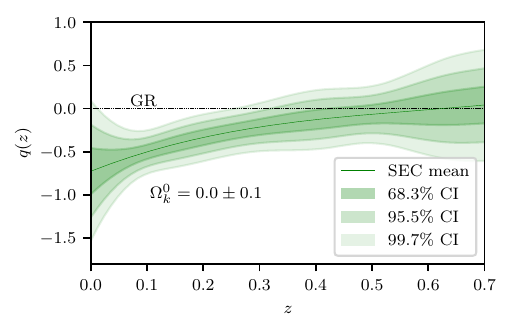}
\caption{\label{fig:sec} The green solid lines are the mean estimates of the SEC, and the green shaded areas are the 68.3\%, 95.5\% and 99.7\% CIs at different redshift values obtained from the reconstructed $q(z)$ by using SNe Ia + BAO + $H(z)$ data. The left painel covers the entire redshift interval studied, whereas the right painel comprehends $z \in [0, \, 0.8]$ highlighting the strongest violated region. This MCMC result was obtained using a Gaussian prior on $\Omega_k^0$ with zero mean and $\sigma = 0.1$.}
\end{figure}

\begin{figure*}
\begin{center}
\includegraphics[scale=0.89]{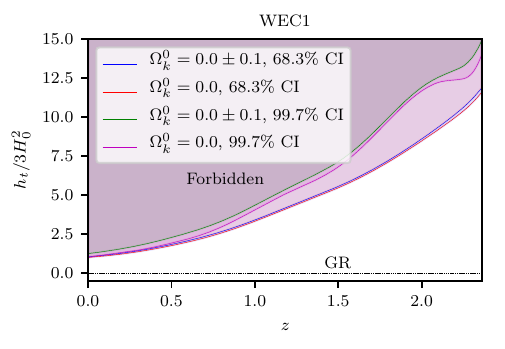}%
\includegraphics[scale=0.89]{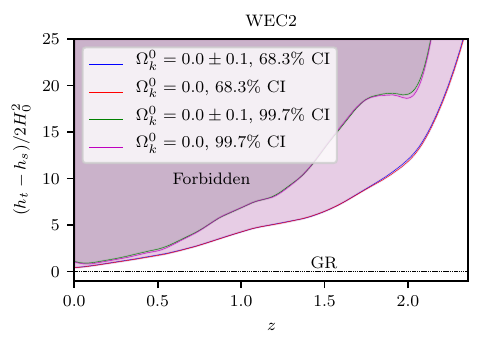}
\end{center}
\begin{center}
\includegraphics[scale=0.89]{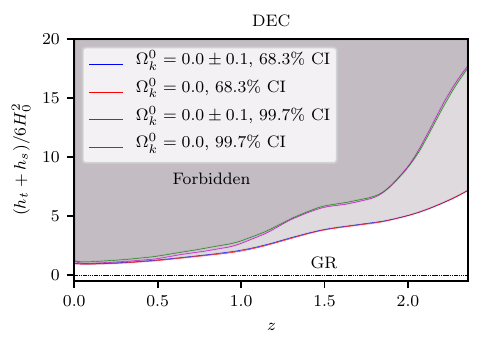}%
\includegraphics[scale=0.89]{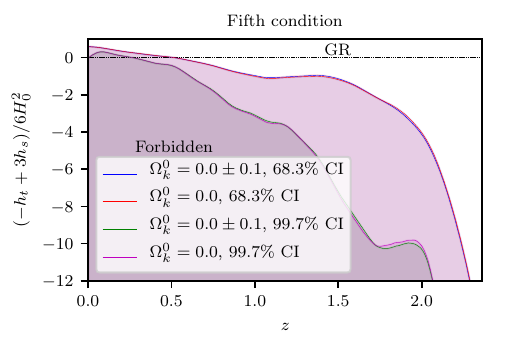}
\end{center}
\caption{\label{fig:ec_eft} The green and pink solid lines are the  $99.7\%$ CIs of the WEC1, WEC2, DEC and fifth condition (FEC) observational bounds. The shaded areas cover the values for which the functions $f(\htt, \hs)$ cannot assume to the respective condition be fulfilled. The dashed lines correspond to the GR fulfillment thresholds.}
\end{figure*}

\begin{figure*}
\begin{center}
\includegraphics[scale=0.89]{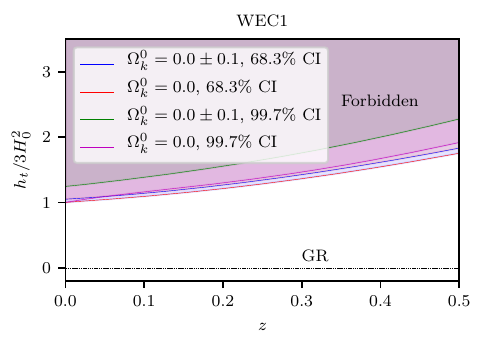}%
\includegraphics[scale=0.89]{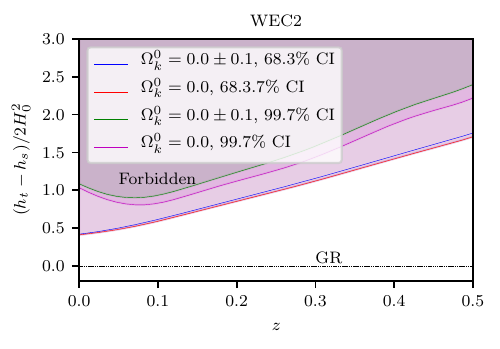}
\end{center}
\begin{center}
\includegraphics[scale=0.89]{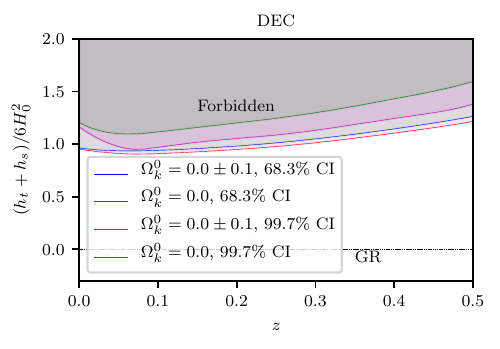}%
\includegraphics[scale=0.89]{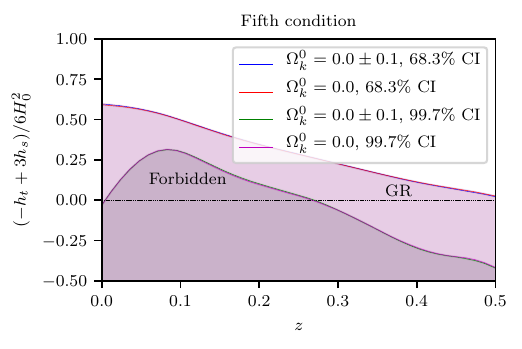}
\end{center}
\caption{\label{fig:ec_eft_zoom} Similar to figure~\ref{fig:ec_eft}. The results are presented in the redshift interval $[0, \, 0.5]$, where the reconstruction of $q(z)$ is better constrained, i.e., presents smaller variances than for higher redshifts.}
\end{figure*}

\begin{figure*}
\includegraphics[scale=0.89]{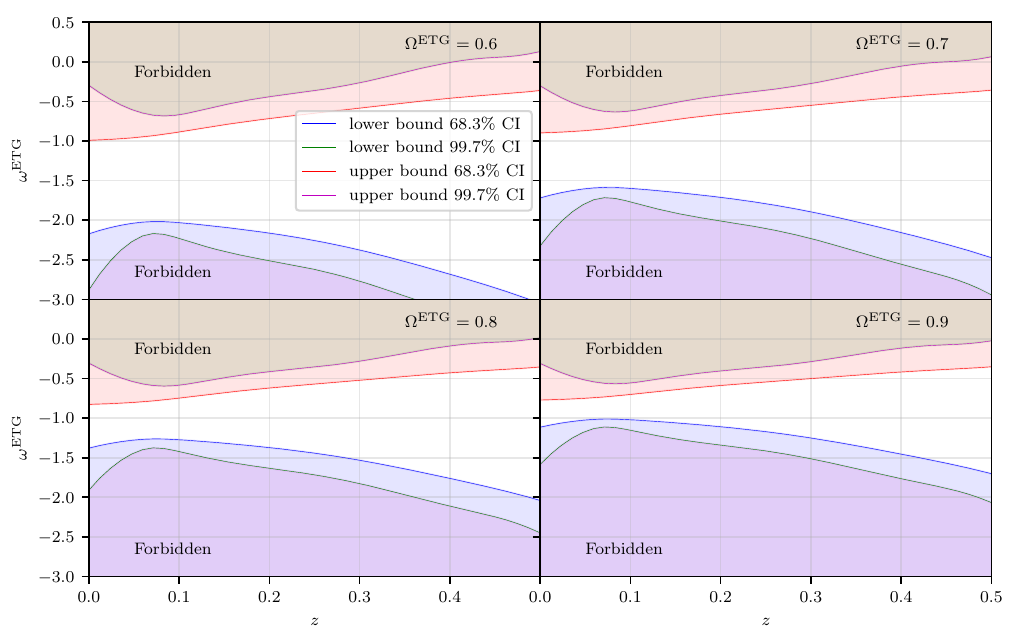}
\caption{\label{fig:wbd} The one and three sigma bounds (68.3\% and 99.7\% CIs) on $\omega^\mathrm{ETG}$ at the redshift interval $[0, 0.5]$ obtained from the reconstructed $q(z)$ by using SNe Ia + BAO + $H(z)$ data. The four plots show the effect of changing the value of $\Omega^\mathrm{ETG}$ on these bounds. These results correspond to the flat case $\Omega_k^0 = 0$.}
\end{figure*}

We first analyze the ECs in the context of GR. From
Eqs.~(\ref{eq:NEC}--\ref{eq:DEC}), we have that the ECs are fulfilled at
any redshift value if they are equal or greater than zero. Note that WEC2
corresponds to the NEC in GR. Figures~\ref{fig:ec_gr} and \ref{fig:sec}
show the MCMC results for the case where we consider a Gaussian prior on
$\Omega_k^0$ with zero mean and scatter 0.10. The green solid lines
represent the ECs means and the shaded areas are the $68.3\%$, $95.5\%$
and $99.7\%$ confidence intervals (CIs) at numerous redshift values in
the range $[0, \; 2.33]$. The results for the other two cases, flat and
$\Omega_k^0 = 0 \pm 0.05$, are pretty similar to this one. One must
be careful when interpreting these plots. Note that each function value, e.g,
$q(z)$, for a fixed value of $z$ has a posterior distribution. The plots
show the mean $\langle q(z) \rangle$ and the $68.3\%$, $95.5\%$ and
$99.7\%$ areas around the mean. Furthermore, these values are obtained by
marginalizing over all other function values and nuisance parameters
($\Omega_k^0, \alpha, \beta, M_1, M_2, H_0, r_d $). Consequently, Figs.~~\ref{fig:ec_gr} -- \ref{fig:wbd} do not show the correlations between different redshifts and must
be interpreted accordingly.

The WEC (Eq.~\eqref{eq:WEC1}, left-lower panel of Fig.~\ref{fig:ec_gr}) is fulfilled in the entire redshift interval, and the DEC (right-lower panel) presents a violation within $99.7\%$ CI just for roughly $z \gtrsim 2.17$ in the three study cases, which is a highly degenerated region due to the few number of data points.

In turn, the upper panel of Fig.~\ref{fig:ec_gr} shows that NEC is
violated within $99.7\%$ CI for $z \lesssim 0.06$, $1.3 \lesssim z
\lesssim 1.8$ and $z \gtrsim 2.08$, where the last is in the degenerated
region of the $q(z)$ [and $H(z)$] reconstruction. We note that these
violation intervals are narrower than those obtained in Refs.~\cite{Lima2008,
Lima2008a} using only SNe Ia data and $z \in [0, \, 1.0]$, where the
violations were present for $z \lesssim 0.1$ and $z \gtrsim 0.8$. Despite of
using a larger SNe Ia sample and BAO and $H(z)$ data, it is worth
mentioning that the reconstruction method used in the present work is
less restrictive than that in \cite{Lima2008, Lima2008a},\footnote{The
current reconstruction is based on cubic splines, whereas
\citet{Lima2008, Lima2008a} used linear splines.} and it was calibrated
such that the bias contributes only with $10\%$ of the total error
budget. The NEC, WEC and DEC results obtained for the flat and
$\Omega_k^0 = 0.0 \pm 0.05$ cases are pretty similar to those presented
in Fig.~\ref{fig:ec_gr}.

The most interesting result concerns SEC, since this is the only EC we
expect to be violated, which is tightly linked with the recent
accelerated expansion of the Universe. Figure~\ref{fig:sec} shows the
reconstructed deceleration function, namely the mean $q(z)$ curve along
with the $68.3\%$, $95.5\%$ and $99.7\%$ CIs. We note (left panel) that
the evidence for SEC violation takes place over the entire redshift
interval. In fact, we obtain SEC's fulfillment within $68.3\%$ CI just
for $0.97 \lesssim z \lesssim 1.02$ (for the other two cases, we have $z
\in [\sim 0.9, \, \sim 1.1]$). In VPL~\cite{Vitenti2015} this fulfillment
was observed in the range $1.84 \lesssim z \lesssim 2.13$. Despite the
reconstruction methodology be the same used here, the differences result
from some distinct and new BAO and $H(z)$ data points
(e.g.,~\cite{Moresco2015, Ata2018, Bautista2017}) used in the analysis.

Contrarily to the new constraints for NEC, SEC violation is stronger than
those obtained in previous work. The right panel of Fig.~\ref{fig:sec}
highlights the redshift interval where we obtain the most restrictive
constraints. For instance, SEC is violated with more than $99.7\%$ CI for
$0.01 \lesssim z \lesssim 0.26$ for the three cases studied, whereas in
VPL~\cite{Vitenti2015} this range was $[\sim 0.02, \, \sim 0.22]$, and
$[\sim 0.1, \, \sim 0.17]$ and $[\sim 0.06, \, \sim 0.2]$  in
Refs.~\cite{Lima2008, Lima2008a}, respectively, using only SNe Ia data.
In short, the most current SNe Ia, BAO and $H(z)$ data strengthens the
evidence of an accelerated expansion of the Universe. For instance, calculating the posterior for $q_\mathrm{min} = \mathrm{min}(q(z))$ for $z\in(0, 0.5)$ we found that $q_\mathrm{min} < 0$ for all points in our sample of the posterior. This means that the probability of finding $q_\mathrm{min} \geq 0$ is smaller than $1 / (5\times10^6)$ (one in the number of posterior sampled points), which translates to at least $5.22\sigma$ confidence level.

As shown above, the fulfillment/violation of the ECs in the GR case can be directly tested since the bounds need just to be compared to constant values (dashed lines in Figs.~\ref{fig:ec_gr} and \ref{fig:sec}). Regarding the class of ETGs considered in this work, this is just valid for NEC and SEC as they have the same form in GR as in these ETGs and, therefore, the results and analyses presented above (upper panel of Figs.~\ref{fig:ec_gr} and \ref{fig:sec}) are also valid for these ETGs. This is to be expected, since these conditions were obtained from the convergence conditions imposed directly on $R_{ab}$ and our reconstruction method output is exactly $R_{ab}$.

On the other hand, the remaining ECs, i.e., WEC1, WEC2 and DEC (Eqs.~\eqref{eq:WEC1}, \eqref{eq:WEC2} and \eqref{eq:DEC}, respectively), and also the FEC [Eq.~\eqref{eq:fifth_cond}] involve not only $q(z)$, $E(z)$ and $\Omega_k^0$, but also the arbitrary functions $\htt(z)$ and $\hs(z)$ of the modified gravity tensor $H_{ab}$. Consequently, instead of checking whether a condition is satisfied or not, our methodology allows one to put constraints on these functions [and their combinations, say $f(\htt, \hs)$].

From the reconstructed $q(z)$ and $E(z)$ curves, we obtain now the observational constraints on the functions $f(\htt, \hs)$ by requiring that WEC1, WEC2, DEC and FEC are fulfilled (see Eqs.~\eqref{eq:WEC1}, \eqref{eq:WEC2}, \eqref{eq:DEC} and \eqref{eq:fifth_cond}). Figure~\ref{fig:ec_eft} shows the $68.3\%$ and $99.7\%$ CIs of the observational bounds obtained for the flat and $\Omega_k^0 = 0 \pm 0.1$ cases. Note that WEC1 (left upper panel), WEC2 (right upper) and DEC (left lower) provide upper bounds on their respective functions $f(\htt, \hs)$ whereas the FEC (right lower panel) gives a lower bound.
The shaded areas in all four panels of Figure~\ref{fig:ec_eft} indicate the values per redshift for which the respective conditions are violated. These \textit{forbidden} areas are wider, i.e., more restrictive as the redshift values decrease due to the larger number of data points. Figure~\ref{fig:ec_eft_zoom} shows the same results for the better constrained region $z \in [0, \, 0.5]$. We see that the GR thresholds, which correspond to the case where $H_{ab} \equiv 0$, i.e., $\htt(z) = \hs(z) = 0$, of WEC1, WEC2, and DEC are fulfilled in the entire redshift interval. On the other hand, a quite different situation is verified for the FEC. The right lower panel of Fig.~\ref{fig:ec_eft_zoom} shows that this condition is violated in GR theory within $99.7\%$ for $0.01 \lesssim z \lesssim 0.26$. Since for GR the FEC is equivalent to SEC, the present result is consistent with that discussed previously (Fig.~\ref{fig:sec}), where SEC violation in GR indicates that the Universe presents an accelerated expansion, which must be driven by a cosmological constant or an exotic fluid with equation of state, for example, of the type $p = w\rho$, with $w < -1/3$.

To go further without choosing a specific ETG, we need to make some
assumptions about $\htt$ and $\hs$. For this it is useful to use the
language of an effective fluid. First we define $\rho^\mathrm{ETG}
\equiv \htt/(8\pi G)$ and $p^\mathrm{ETG} \equiv - \hs/(8\pi G)$. Rewriting DEC and FEC using these variables, we can impose upper and lower
bounds on $\Omega^\mathrm{ETG}\omega^\mathrm{ETG}$, i.e., 
\begin{align}
\Omega^\mathrm{ETG}\omega^\mathrm{ETG} &\geq -2\frac{[2 - q(z)]E(z)^2 - 2\Omega_k^0(1+z)^2}{3} \nonumber \\&+ \Omega^\mathrm{ETG}, \\
\Omega^\mathrm{ETG}\omega^\mathrm{ETG} &\leq 2\frac{q(z)E(z)^2}{3} - \frac{\Omega^\mathrm{ETG}}{3},
\end{align}
where $\omega^\mathrm{ETG}
\equiv p^\mathrm{ETG} / \rho^\mathrm{ETG}$ and $\Omega^\mathrm{ETG} \equiv 8\pi G \rho^\mathrm{ETG}/(3H_0^2)$. If the DE was explained by a dark fluid instead of an ETG, the same bounds above would be obtained. Nonetheless, in this case the FEC requires again that all fluids but the DE fluid  
satisfy the SEC. In addition, the upper bound (derived from DEC) requires a similar assumption, all fluids but the DE one must satisfy the DEC. It is interesting to note that the last mentioned condition is not necessary in the ETG case since, by definition, DEC already imposes conditions only on the ordinary matter.

Assuming that $\Omega^\mathrm{ETG} > 0$, we divide the bounds
above by $\Omega^\mathrm{ETG}$ to obtain bounds on $\omega^\mathrm{ETG}$. In
Fig.~\ref{fig:wbd} we plot these bounds considering $\Omega^\mathrm{ETG}$
constant. Even though we do not expect, a priori, that
$\Omega^\mathrm{ETG}$ is constant, these bounds plotted with different values of $\Omega^\mathrm{ETG}$ show approximately the bounds one would obtain if $\Omega^\mathrm{ETG}$ smoothly evolved inside the chosen redshift values. This
amounts to show how these bounds can impose constraints on the behavior of the ETG. For instance, in Fig.~\ref{fig:wbd} we note that the lower the $\Omega^\mathrm{ETG}$ value, the higher the evidence to obtain a phantom-like behavior. Naturally, the
value of $\Omega^\mathrm{ETG}$ today must be closer to the one estimated by
current data (Ref.~\cite{Planck2015}, for example, gives
$\Omega^\mathrm{ETG}_0 \approx 0.7$). Nevertheless, for $z > 0$ the
evolution of $\Omega^\mathrm{ETG}$ can take it to different values
depending on the specific dynamic of the ETG and, therefore, these bounds show the restrictions depending on how this function evolves in time.

\section{Conclusion}
\label{conclusions}

Numerous propositions of modified gravity theories and also of the DE equation of state in the context of GR have been introduced and discussed in the last two decades to explain the recent accelerated expansion of the Universe. Therefore, one major task in cosmology is  to constrain these models and also scrutinize their viability using  observational data.

In this work we introduced a methodology to obtain observational bounds on a class of ETGs and also on the parameter space of a specific theory. This approach consists in requiring the fulfillment of the ECs, from which we obtain the theoretical bounds for the ETGs' functions and/or paramaters.

We derived the ECs for ETGs for which matter is minimally coupled to geometry. Then, assuming a homogeneous and isotropic metric, we wrote these ECs in terms of the functions $q(z)$ and $E(z)$, and the parameter $\Omega_k^0$. We also considered a fifth condition (called, for simplicity, FEC) stating that in the context of ETG there is no need to add fluids, other than the ordinary and dark matter, to explain the accelerated expansion of the Universe. Using the VPL model-independent reconstruction method \cite{Vitenti2015} and SNe Ia, BAO, H(z) data, we obtained observational bounds on combinations of $\htt(z)$ and $\hs(z)$ as given in Eqs.~\eqref{eq:WEC1}, \eqref{eq:WEC2}, \eqref{eq:DEC} and \eqref{eq:fifth_cond}.

We first studied the ECs in GR. We verified that WEC and DEC are fulfilled in the redshift interval $[0, \, 2.33]$. NEC is violated at very low redshifts $z \lesssim 0.06$ (and also at higher $z$ values where the variance of the estimated curve is big, see Sect.~\ref{sec:results}). It is worth emphasizing that NEC violation is weaker than those obtained in \cite{Lima2008,Lima2008a}. On the other hand, the evidence of SEC violation obtained in the present work is stronger than the previous ones \cite{Lima2008,Lima2008a,Vitenti2015}. In particular, there is an indication bigger than $5.22\sigma$ CI of the recent accelerated expansion of the Universe. 

We also obtained the allowed/forbidden values for the WEC1, WEC2, DEC
and FEC functions of $\htt(z)$ and $\hs(z)$,
as showed in Figs.\ref{fig:ec_eft} and \ref{fig:ec_eft_zoom}. Particularly, we have that GR violates the FEC within $99.7\%$ CI for  $z \in [0.01, \, 0.26]$ which is equivalent to the violation of SEC. This reinforces our main idea that if the FEC if not fulfilled, than the theory requires the introduction of a DE fluid to explain the accelerated expansion of the Universe.

Notice that the present study considered a general class of ETGs and, therefore, we just obtained lower and upper bounds for different functions of $\htt(z)$ and $\hs(z)$. However, applying this approach for a specific model, one will be able to obtain upper and lower bounds for a given function or parameters. 

At last, in this regard we provided an example of possible results. By assuming a positive $\Omega^\mathrm{ETG}$, we obtained bounds on the effective ETG equation of state as shown in Fig.~\ref{fig:wbd}. This proves to be potentially useful to determine the behavior of the ETG in the reconstructed redshift interval, from a phantom like behavior, $\omega^\mathrm{ETG} < -1$, or the opposite direction, $\omega^\mathrm{ETG} > -1$, depending on how $\Omega^\mathrm{ETG}$ evolves in time. Therefore, we consider that this method is a useful tool to constrain the parameter spaces of different ETGs. For instance, in Ref.~\cite{Alves2017} we considered an ETG whose modified gravity term, i.e., the tensor $H_{ab}$, acts as a cosmological constant. In this context, we also studied two bimetric massive gravity theories putting constraints on their parameter space. 

\section*{ACKNOWLEDGMENTS}

MPL thanks the financial support from Labex ENIGMASS. SDPV acknowledges the support from BELSPO non-EU postdoctoral fellowship.
MESA and JCNA would like to thank the Brazilian agency FAPESP for financial support under the thematic project \# 2013/26258-4. JCNA thanks also the Brazilian agency CNPq (Grant No. 307217/2016-7) for the financial support. FCC was supported by CNPq/FAPERN/PRONEM. This research was performed using the Mesu-UV supercomputer of the Pierre \& Marie Curie University | France (UPMC) and the computer cluster of the State University of the Rio Grande do Norte (UERN) | Brazil. This study was financed in part by the Coordenação de Aperfeiçoamento de Pessoal de Nível Superior - Brasil (CAPES) - Finance Code 001.

\bibliography{references}
\bibliographystyle{apsrev}

\end{document}